\pdfoutput=1 %arXiv admin added this line
\documentclass[conference]{IEEEtran}

\ifCLASSINFOpdf
\else
\fi

\usepackage{amssymb}
\usepackage{graphicx}
\usepackage[usenames,dvipsnames]{color}

\usepackage{url}

\usepackage{amsmath}
\usepackage{textcomp} % \textlbrackdbl, \textrbrackdbl
\usepackage{xspace} 
\usepackage{bbding} % check marks
\usepackage{pifont} % arrows
\usepackage{stmaryrd} % arrows
\usepackage{mdwlist} %list
\usepackage[cp1252]{inputenc}

\usepackage{focus}

\begin{document}
\newcommand{\autofocusString}{\textsc{AutoFocus}}
\newcommand{\autofocus}{\autofocusString\xspace}
\newcommand{\af}{\autofocus}
\newcommand{\autofocusIII}{\autofocus~3\xspace}
\newcommand{\afIII}{\autofocusIII}
\newcommand{\aft}{\autofocusIII}

\newcommand{\idep}{\mathbb{I}^\mathcal{D}}
\newcommand{\odep}{\mathbb{O}^\mathcal{D}}
\newcommand{\instreams}{\mathbb{I}}
\newcommand{\outstreams}{\mathbb{O}}

\title{ 
Towards Logical Architecture and Formal Analysis of Dependencies Between Services 
}

\author{\IEEEauthorblockN{
Maria Spichkova  
and
Heinrich Schmidt   } 
\IEEEauthorblockA{ 
Computer Science and IT, RMIT University, Melbourne, AUSTRALIA
\\ 
Email: \{Maria.Spichkova,Heinz.Schmidt\}@rmit.edu.au
}
}
\maketitle

\begin{abstract} 
This paper presents a formal approach to modelling and analysis of data and control flow dependencies between services
 within remotely deployed distributed systems of services. 
Our work aims at elaborating for a concrete system, which parts of 
the system (or system model) are necessary to check a given property. 
The approach allows services decomposition oriented towards  efficient checking of system 
properties as well as analysis of dependencies within a system.  
\end{abstract}
 
 \begin{IEEEkeywords}
formal methods; static analysis; dependencies between services; decomposition; verification
\end{IEEEkeywords}
 
\section{Introduction}
 
This approach originated from the analysis of two case studies from automotive area, which were developed together with industrial partners within DenTUM and Verisoft-XT\footnote{Verisoft-XT project -- \url{http://www.verisoftxt.de/}} projects. 
The first case study~\cite{dentum_tb}, developing an Adaptive Cruise Control   system with Pre-Crash Safety functionality,  was motivated and supported by DENSO Corporation, the second  case study \cite{spichkova_tb_decomp}, developing a Cruise Control System with focus on system architecture and verification, was supported by Robert Bosch GmbH.
One of the essential questions we have investigated during this analysis was  
which part of the system functionality do we need to analyse to check  a certain property in sense of monitoring, testing or formal verification. 
In most cases, we don't need complete information about the system as a whole to analyse certain aspects or to check certain properties. 
The problem is how to find which  local information (at the level of one or more components or subsystems) is sufficient for this check, 
and how to solve this problem on a formal level to allow the automatisation of the analysis.

The reason for this question has a very practical ground: any additional information about the system can make the whole process
slower, more expensive or even infeasible,  especially if we are speaking about verification. 
In the case of model checking this can lead to the state explosion problem. 
In the case of theorem proving this can lead to stack overflow or excessive or
prohibitive amounts of time needed by verification engineers to
complete the task.

On the logical level, this problem can be reformulated as follows:
What is the minimal part of model  needed to check a specific property? 
We suggest  an approach focusing on data and control flow 
dependencies between services. % 
Dependencies' analysis results in a
decomposition that gives rise to a logical system architecture, which is the most appropriate for the
case of remote monitoring, testing and/or verification.

In the case where the check of properties depends on information in one
location, obviously we can check the property without remote
connection and without sending  system-wide data for
inclusion in the monitoring, testing, or verification process. 
 In the case of remote
connection, an additional problem arises through the need of
transferring large amounts of data, for example from sensor
instruments or automated test generators and model checkers. 
Thus, it is crucial to analyse regarding  to efficiency, which properties (or their parts) should be checked locally, 
and which should be sent to the cloud. % % 
Therefore, an appropriate model covering all these aspects of remotely deployed distributed systems is needed 
to fulfil the corresponding constraints and to make decisions on (logical) service-oriented system architecture at least semi-automatically.

\emph{Contributions:} 
In this paper, we present a formal approach that allows 
 modelling of data  and control flow dependencies between services. % 
 The aim of the approach is to allow % 
  decomposition oriented towards efficient checking of system
properties, also taking into account such characteristics as performance, worst-case execution time, reliability, etc. 
Another contribution of this approach is a semi-automatic support of these ideas on verification level. 
Applying ideas introduced in this paper within specification and proof methodology \cite{FocusStreamsCaseStudies-AFP}, 
and taking into account human factor analysis~\cite{Spichkova2013HFFM,hffm_spichkova}, we  
obtain concise and at the same time readable specifications. 
To support this approach on verification level, we have built the corresponding set of theories in Higher-Order Logic. 
It allows to check dependency relations  using an interactive % 
theorem prover Isabelle/HOL~\cite{npw}. 
To discharge proof goals % 
automatically, we apply Isabelle's 
component Sledgehammer~\cite{Sledgehammer} that employs resolution based  % 
first-order automatic theorem provers (ATPs) 
and satisfiability modulo theories (SMT) 
solvers.

%

%==================================================
\section{Inter-/Intraservice Dependencies}
\label{sec:communication}

In this paper, we follow the definition of services introduced in \cite{broy_janus}, where a service $S$ is defined as a partial function 
from input streams $\instreams(S)$ to output streams $\outstreams(S)$. 
Thus, the behaviour of a service can be defined partially (only for a subset of its possible inputs), in contrast to the totality of a component interface behaviour. 
In our approach, we need only an abstract definition of a service: we focus on the analysis of dependencies within and between services, where the behaviour of a
(sub)services can be specified using different languages.

While modelling  communication between services  
on a certain abstraction level $L$ (i.e. level of refinement/decom\-po\-sition),
  we specify the  
  dependencies by the  function  
\newpage
\[
Sources^{L}: CSet^{L} \to  (CSet^{L}~set)
\]
$CSet^{L}$ denotes here the set of services at the level $L$. 
$Sources^{L}$ returns for any service   $A$ the corresponding (possibly
empty) set of services   
 that are the sources for the input streams of $A$. 
More precisely, all the dependencies can be divided into the direct and indirect ones. 
As a \emph{stream} we understand here both data and control flows. 
On the level of logical architecture, we can see a stream as an abstract \emph{channel}.

Similar to the function $Sources^{L}$, we define direct dependencies by the function 
\[
DSources^{L}: CSet^{L} \to  (CSet^{L}~set)
\]
For example, 
$C_1 \in DSources^{L}(C_2)$ simply means that at least one of the output channels of  $C_1$ 
is directly connected to some of input channels of   $C_2$.
In more complicated situations, we need
to reason about the (transient)  dependencies within many or all of the
services of the system.

The direct  sources for $C$ can be defined as follows: 
\[
DSources^{L}(C) = 
\{  S \mid  \exists x \in \instreams(C) \wedge x \in \outstreams(S) 
\} 
\]
If we do not have to take into account the indirect  dependencies, 
the    function  $Sources^{L}$ is defined to be equal to the  function  $DSources^{L}$, otherwise 
 we define this function recursively over the set of services using the following fixed point construction:
\[
\begin{array}{l}
Sources^{L}(C) = \\
DSources^{L}(C)\ \cup\  \bigcup_{S \in  DSources^{L}(C)} \{  S_1 \mid S_1 \in  Sources^{L}(S) \} 
\end{array}
\]
The functions $DSources^{L}(C) $ and $Sources^{L}(C)$ have a number of crucial properties. 
For example:

\noindent
\emph{(1)} %Lemma 1:
 If a service $X$ does not belong to the $DSources^{L}$ set of any service $C$ of the system, it also does not belong  
to the $Sources^{L}$ set of any service $C$:
\[
\begin{array}{l}
\forall C \in CSet^{L}.\ X \notin DSources^{L}(C) \Rightarrow\\
 \forall C \in CSet^{L}.\ X \notin Sources^{L}(C)
\end{array}
\]
Thus, if $C$ belongs to the $Sources^{L}$ set of a service $S$, then exist some  service $Z$ s.t. $C$ belongs to its $DSources^{L}$ set 
(as a special case, $Z$ could also be equal to $S$):
\[
\begin{array}{l}
C \in Sources^{L}(S)  \Rightarrow \exists Z.\ C  \in DSources^{L}(Z)
\end{array}
\]
This also imply that if $Sources^{L}$ set of a  service $C$ is empty, its $DSources^{L}$ set is empty too, and vice versa:
\[
\begin{array}{l}
Sources^{L}(C) = \emptyset  \Leftrightarrow   DSources^{L}(C) = \emptyset
\end{array}
\]
\emph{(2)} %Lemma 2 (Transitivity):
Transitivity: If a  service $C$ belongs to the $Sources^{L}$ set of a  service $S$, which itself belongs to the $Sources^{L}$ set of another  service $Z$, 
then the  service $C$ also belongs to the $Sources^{L}$ set of a  service $Z$.
\[
\begin{array}{l}
C \in Sources^{L}(S)  \wedge S \in Sources^{L}(Z) \Rightarrow C  \in Sources^{L}(Z)
\end{array}
\]
\emph{(3)} If we have mutual dependencies between  services, then their $Sources^{L}$ sets are equal and contain these  services themselves 
($XS$ and $ZS$ denote here some sets over $CSet^{L}$): 
\[
\begin{array}{l}
Sources^{L}(C) = (XS \cup Sources^{L}(S)) \wedge \\
Sources^{L}(S) = (ZS \cup Sources^{L}(C))\\
 \Rightarrow Sources^{L}(C) = Sources^{L}(S) = XS \cup ZS \cup \{ C, S\}
\end{array}
\]

In general, values of an output channel $y \in \outstreams(C)$ of a  service $C$ do not necessarily depend on the values of all its input streams. 
This means that an optimisation of system/ services' architecture may be needed in order to localise these dependencies. 
To express any restrictions we use the following notation: $\idep(C, y)$ denotes the subset of $\instreams(C)$ that $y$ depends upon. 
There are three possible cases to consider:
\begin{itemize}
\item 
$y$ depends on  all input streams of  $C$: %\\
 $\idep(C, y) = \instreams(C)$;
\item
$y$ depends on some input streams of $C$: %\\
$\idep(C, y) \subset \instreams(C)$;
\item 
$y$ is independent of any input stream of $C$, i.e.,\\
 $\idep(C, y) = \emptyset \not = \instreams(C)$.
\end{itemize}

While determining these sets, we should take into account not only the direct
dependence between input/output values, but also the dependence via
local variables of the  service. 
For example, let $C$ be a service with a local variable $st$, representing its current state, 
and an output
channel $y$, which depends only on the value of $st$.
If $st$ is updated depending to the input messages
the  service receives via the input channel $x$, then $\idep(C, y) = \{ x \}$.    
To be more precise, we mark this special case by an upper
index $\idep(C, y) = \{ x^{(st)} \}$ indicating that $x$ influences $st$, and $y$ via $st$.

\subsection{Elementary Services}
\label{sec:elementary}

Based on the definition above, we can decompose services to have for each  output channel 
the minimal sub service computing the corresponding results (we call them \emph{elementary  services}). 
An elementary  service either 
\begin{itemize}
\item
should have a single output channel (in this case this  service can have no local variables), 
or 
\item
all its output channels are correlated, i.e. mutually depend on the same local variable(s).
\end{itemize}
Let $C$ be a service with $m$ input channels $x_{1}, \dots, x_{m}$ 
and $n$ output channels $y_{1}, \dots, y_{n}$ as well as $k$ local variables $l_{1}, \dots, l_{k}$. 
For each output channel $y_{i}$, $1 \le i \le n$ we check the corresponding set $\idep(C, y_{i})$:\\
(1)
If $\idep(C, y_{i})$ contains only direct dependencies from some input channels, i.e. $\idep(C, y_{i}) \subseteq \instreams(C)$, 
we remove the corresponding computations from $C$ to a single subservice $C_{i}$, 
which has a single output channel $y_{i}$.\\
(2)
If $\idep(C, y_{i})$ contains also dependencies via some local variables, then we 
should check which other output channels depend from these variables, because 
the output channels depending from the same local variables will belong to the same subservice of $C$, otherwise we will need duplicated computation for these variables. 
Thus, we proceed as follows: 
	\begin{itemize}
	\item
	If $y_{i}$ is the only output channel depending from these variables, 
	we remove the corresponding computations from $C$ to a single subservice $C_{i}$, 
        which has a single output channel $y_{i}$.
        \item 
        If there are other $jm$ output channels $y_{j1}, \dots, y_{jm}$ depending from these variables, 
        we remove the corresponding computations from $C$ to a single subservice $C_{i}$, 
        which has $jm+1$ output channels $y_{i}, y_{j1}, \dots, y_{jm}$.
	\end{itemize} 
For all three cases, $\idep(C, y_{i})$ becomes a set of input channels of the new subservice $C_{i}$.  
If after this
decomposition a single service is too complex, we can apply the
decomposition strategy presented in~\cite{spichkova2011decomp}.

%--------------
\subsection{Running Example}
\label{sec:example}

\begin{figure}[ht!]
 \centering 
   \includegraphics[width=9cm, natwidth=489,natheight=159]{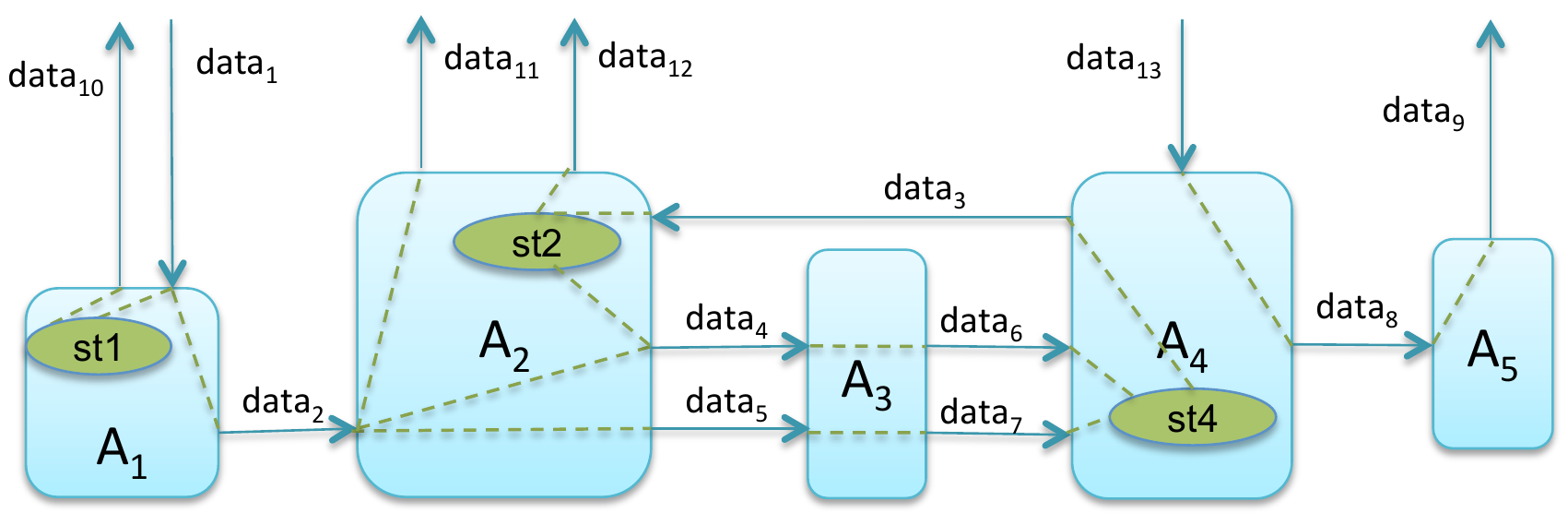} 
    \caption{System  $S$: Dependencies and $\idep$ sets }
    \label{fig:example_comm1}
\end{figure}

Let us illustrate the presented ideas by an example: %First of all, 
we show how each service can be decomposed to optimise
the  dependencies within each single service, and after that we optimise the architecture of the whole system. 
Given a system
$S$ (cf.\ also Fig.~\ref{fig:example_comm1}) consisting of five services, where the set $CSet$ on the level $L_{0}$ is
defined by $\{A_1, \dots, A_5\}$. 
The sets $\idep$ of data and control flow dependencies between the services are shown in Table~\ref{table:dep_exampleC}. 
We represent the dependencies graphically using dashed lines over the service box.

\begin{figure}[ht!]
\centering
   \includegraphics[width=9cm, natwidth=497,natheight=259]{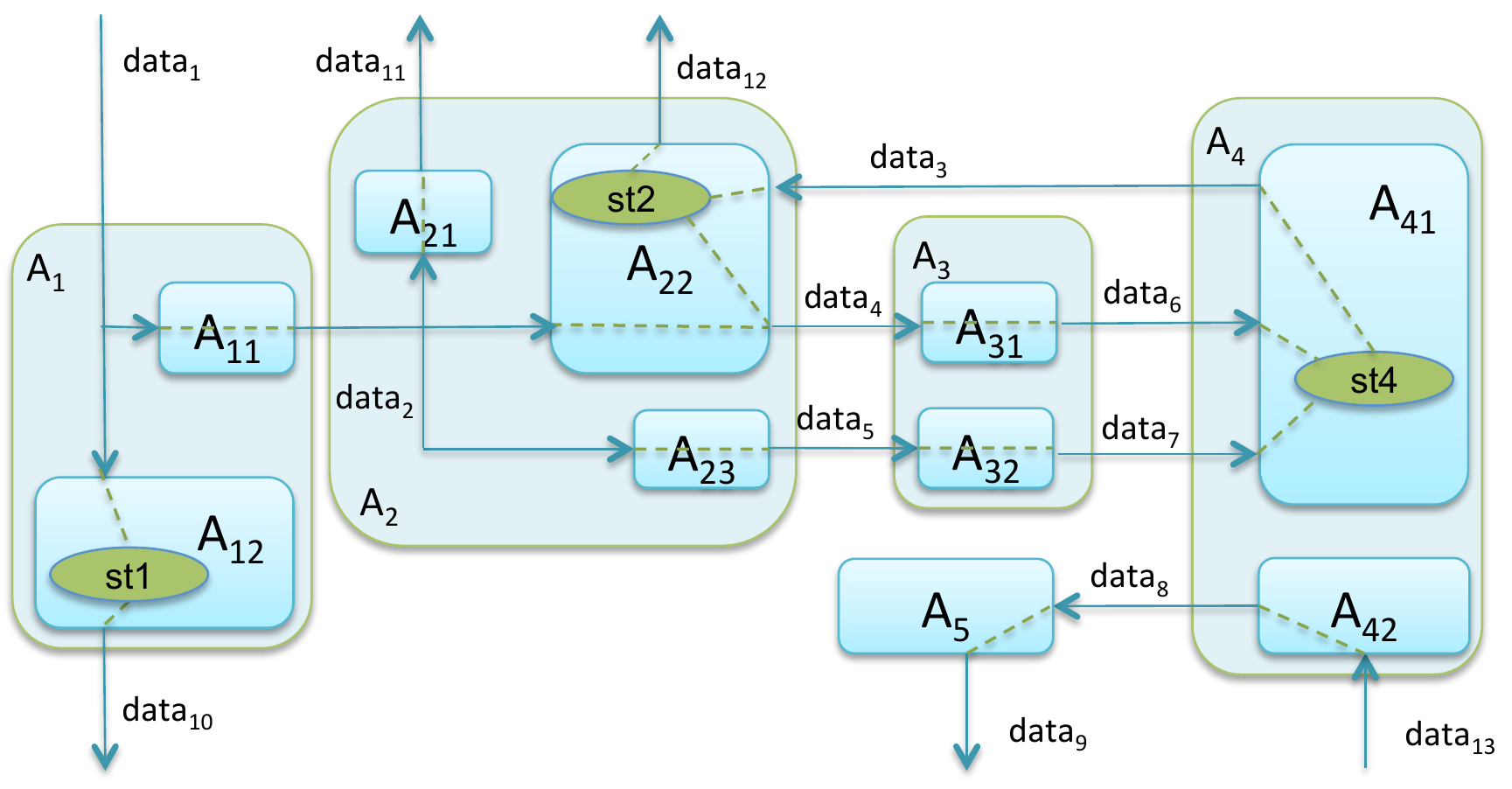}
    \caption{  
    Services' decomposition (level $L_{1}$)}
    \label{fig:example_comm2}
\end{figure}

\noindent
Now we can decompose the system's services according to the given  $\idep$ specification. This results into the next abstraction level $L_{1}$ of logical architecture (cf. Fig.~\ref{fig:example_comm2}), on which all services are elementary:
\begin{itemize}
\item[$A_{1}$] has two output channels, 
$data_{2}$ depending on the input channel $data_{1}$ directly and $data_{10}$ depending on the same input channel but via a local variable $st1$. 
Therefore, it should be decomposed in two subservices: 
$A_{11}$ with input $data_{1}$ and output $data_{2}$, and $A_{12}$ with input $data_{1}$ and output $data_{10}$.
\item[$A_{2}$] has four output channels: 
$data_{5}$ and $data_{11}$ depend on the input channel $data_{2}$ directly, the corresponding computations should be removed to two subservice, $A_{23}$ and $A_{21}$, respectively.
$data_{12}$ depends on the input channel $data_{3}$ via local variable $st2$, but there is another output channel $data_{4}$ depending on this local variable, therefore, computations for these two outputs should belong to the same subservice $A_{22}$. Thus, $A_{2}$ should be decomposed in three subservices.   
\item[$A_{3}$] has two output channels, 
$data_{6}$ depending on the input channel $data_{4}$ directly and $data_{7}$ depending on the input channel $data_{5}$ directly. 
Therefore, it should be decomposed in two subservices.   
\item[$A_{4}$] has two output channels, $data_{8}$ and $data_{3}$. 
It should be decomposed in two subservices, because
$data_{8}$ depends on the input channel $data_{13}$ directly and 
$data_{3}$ depends on the input channels $data_{6}$ and $data_{7}$ via local variable $st4$. 
\item[$A_{5}$] has only one output channel, which means that no decomposition on this level is needed.
\end{itemize} 

\begin{table}[ht!]
\caption{Dependencies within the system  $S$ }
 {\footnotesize 
\begin{center}
\begin{tabular}{|c | c | c |  l  |}
 \hline
 &	  $DSources^{L_{0}}$  	& $Sources^{L_{0}}$ 			& ~ $\idep$ 
 \\
 \hline
 $A_1$	 & $\emptyset$			& $\emptyset$				&  ~ $data_2:\  \{data_1\}$ \\
 
 			&&& ~ $data_{10}:\  \{data_1^{st1}\}$  \\
  \hline
  $A_2$	 & $\{ A_1,  A_4\}$  		& $\{ A_1, A_2, A_3, A_4\}$    	& ~ $data_4:\  \{data_2, data_{3}^{st2}\}$ 	\\
  &&& ~ $data_5:\  \{data_2\}$  \\ 
  		 &					&						& ~ $data_{11}:\  \{data_2\}$			\\
		 &&& ~  $data_{12}:\  \{data_{3}^{st2}\}$ \\
  \hline
 $A_3$     & $\{ A_2 \}$  		& $\{ A_1, A_2, A_3, A_4\}$ 		& ~ $data_6:\  \{data_4\}$ 			\\
 &&	& ~  $data_7:\  \{data_5\}$  \\ 
 \hline
  $A_4$  	& $\{ A_3\}$  		& $\{ A_1, A_2, A_3, A_4\}$  		& ~ $data_3:\  \{data_6^{st4}, data_{7}^{st4}\}$  
  \\
  &&&~  $data_8:\  \{data_{13}\}$  \\ 
 \hline
  $A_5$   &  $\{ A_4\}$		&  $\{ A_1, A_2, A_3, A_4\}$		& ~ $data_9:\  \{data_8\}$	 
 \\
  \hline
\end{tabular}
\end{center}
}
\label{table:dep_exampleC}
\end{table}%

\section{Reliability Analysis and Tracing} 
\label{sec:reliability}

The functions $Sources^{L}$ and $\idep$  allow us to trace back which parts of the system provide information to a certain service. This provides a basis for identifying of elementary services and the corresponding optimisation of the logical architecture.
To trace which services are affected by the results produced by the service, we introduce another two functions,  
$\odep$  and $Acc^{L}$. 
For any service $C$, the function $\odep$ (dual to $\idep$) returns the
corresponding set $\odep(C,x)$ of output channels depending on input
$x$. 
This help us to solve following problems: 
\begin{itemize}
\item
if there are some changes in the specification, properties, constraints, etc.\ for $x$, 
we can trace which other channels can be affected by these changes;
\item
having for each output channel  the minimal subservice computing the corresponding results, 
we can check the critical paths in the system by allocating to each subservice /output channel the corresponding 
worst case execution time (WCET). 
For this purpose we can use representation of the system architecture as a directed graph (cf. Section~\ref{sec:optim}).
\end{itemize} 
If the input part of the service's interface is specified correctly in the
sense that the service does not have any ``unused'' input channels, the following relation
will hold: 
\[
\forall x \in \instreams(C). ~ \odep(C, x) \neq
\emptyset.
\]
On each abstraction level $L$ of logical architecture, we can define a function 
$
Acc^{L}: CSet^{L} \to  (CSet^{L}~set)
$.
For any service (name) $A$ the function $Acc^{L}$ returns the corresponding (possibly
empty) set of services (names) $B_1, \dots, B_{AN}$ that are the
acceptors for the output  streams of $A$. 
This function  dual to the function $Sources^{L}$:
\[
\begin{array}{l}
x \in Acc^{L}(y) \iff  y \in Sources^{L}(x)
\end{array}
\]
This model allows us to analyse the influence of a service's failure on the functionality of the overall system:
if a service $C$ fails, the set of affected function will be $Acc^{L}(C)$.
For example,  for the system from  Fig.~\ref{fig:example_comm2},  on the abstraction level $L_{1}$ 
the services $A_{12}$, $A_{21}$, $A_{5}$, 
have no acceptors, where the set of acceptors of the service $A_{11}$ is 
$\{ A_{21}, A_{22}, A_{23}, A_{31}, A_{32}, A_{41}\}$. 
Thus, failure of the service $A_{12}$ causes wrong behaviour only of $A_{12}$  itself, where 
failure of the service $A_{12}$ has influence on the results of seven services (including $A_{12}$).
On this basis, we can mark each service $C$ by a number of services which are affected by $C$ (and,  respectively, by its failure). Thus, the \emph{impact number} of the service $A_{12}$  will be 1 (the minimal value), where the impact number of $A_{11}$ will be 7.
Moreover, on this basis we can apply results of our previous work on 
efficient hazard and impact analysis for automotive mechatronics systems \cite{hazardTB}, which was done in collaboration with industrial partners from ITK Engineering AG.

The representation of dependencies between services by a directed graph (like on Fig.~\ref{fig:graph})
 allows also to find the worst case execution time (WCET) needed for the concrete output based on the   
WCETs of the system's services, as well as analyse the influence of a service's failure on overall system.
For example, from Fig.~\ref{fig:graph} it is easy to see that for the outputs 
of the services $A_{11}$ and $A_{12}$ the worst case computation time is equal to the WCETs of these services, for the output
of the services $A_{21}$  the worst case computation time is equal 
to the sum of WCETs of $A_{11}$ and $A_{21}$, etc.
\section{Strongly Connected Services} %  
\label{sec:optim}

After the decomposition discussed in the previous section, we obtain a  (flat) architecture of system. 
The main feature of this architecture is that  each output channel (within the system) 
belongs the minimal subservice of a system computing the corresponding results. 

We represent this (flat) architecture  as a directed graph % 
and apply one of the existing distributed algorithms for the 
decomposition into its   strongly connected services,  e.g. FB~\cite{idetifyingSSCs}, OBF~\cite{OBF}, or the colouring algorithm~\cite{Orzan04ondistributed}.
For our goals, we extend them by a preliminary simplification of the graph (cf. below). 
This optimisation is algorithm independent and is also applicable for the case another decomposition is chosen.

Let us introduce some basic terms and definitions to explain the main ideas of the decomposition we apply in our approach. 
A \emph{directed graph} $G$ is a pair $(V, E)$, where $V$ is a set of vertices, and $E \subseteq V \times V$ is a set of directed edges. 
If $(u,v) \in E$, then $v$ is called an \emph{successor} of $u$, and $u$ is called \emph{predecessor} of $v$. 
A vertex $t$ is called \emph{reachable} from vertex $x$ if $(x,t) \in E^*$, where $E^*$ is a transitive and reflexive closure of $E$. 
If $x_k$ is reachable from $x_0$, then there is a sequence of vertices $x_0, \dots, x_k$ (called \emph{path}), such that 
for all $0 \le i < k$ $(x_i, x_{i+1}) \in E$.  

A set of vertices $SC \subseteq V$ is \emph{strongly connected} if for any vertices $u, v \in CS$ holds that $v$ is reachable from $u$. 
A \emph{strongly connected service (SCS)} is a maximal straggly connected set of vertices, i.e. after
extension of this set by any additional vertex the set is no more strongly connected. 
An SCS is \emph{trivial} if it consists of a single vertex. We call an SCS \emph{leading (terminating) trivial} if it consists of a single vertex that has no predecessors (successors). 

\begin{figure}[ht!]
  \begin{center}
   \includegraphics[width=8cm, natwidth=314,natheight=167]{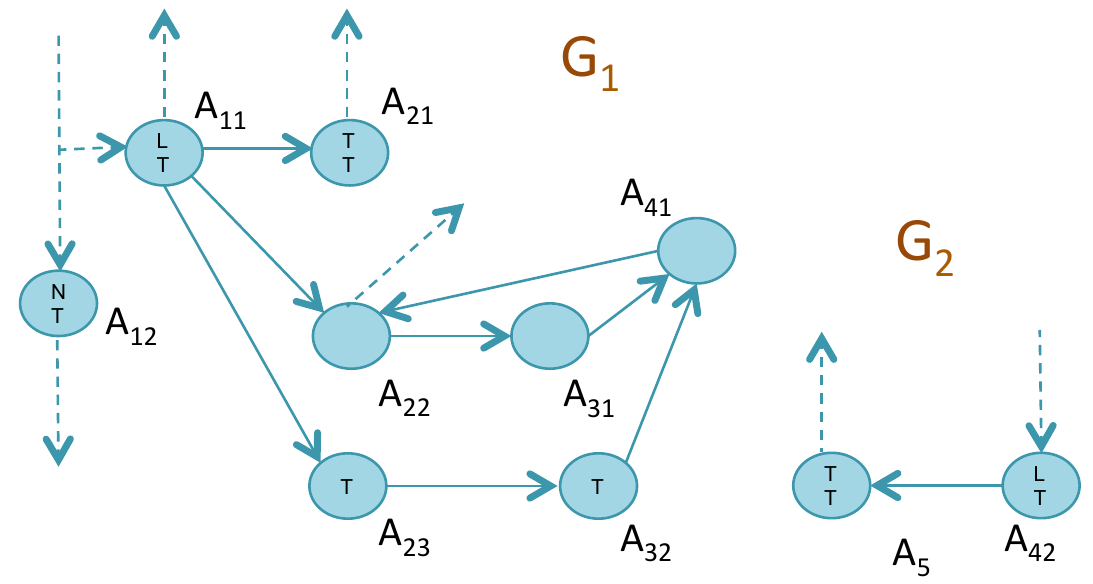}
    \caption{% 
    System $S$  (level $L_{1}$): 
    detection of the strongly connected services }
    \label{fig:graph}
  \end{center}
\end{figure}

In our representation of a flat architecture  as a directed graph, services become vertices and channels become edges. 
We distinguish here
\begin{itemize}
\item channels that are system's input/output  
(within the graph, we represent them by dashed edges, cf. also Fig.~\ref{fig:graph}, because they are less important on this level of (de)composition -- we do not need to take them into account by identifying the SCSs), 
and 
\item
channels that are local for the system, i.e. representing dependencies between services.
\end{itemize}
Vertices that are leading trivial SCSs are labeled by LT, vertices that are terminating trivial SCSs are labeled by TT, and
vertices that are \emph{commonly trivial} SCSs (trivial, but neither leading nor terminating) are simply labeled by T. 
In addition, we define another special king of trivial SCSs: services that has neither successors, no predecessors
 (this means that  all their input/output channels are system's input/output channels).  
 We call them \emph{disconnected trivial} SCSs and label by DT.

Two vertices $u$ and $v$  of an undirected graph $G$ are  \emph{connected} if there is a path  from $u$ to $v$ (for the case of directed graph we need to ignore the orientation of its edges);
otherwise, they are   \emph{disconnected}. 
A graph $G$ is \emph{connected} if every pair of  vertices in $G$ is connected; otherwise, it is called \emph{disconnected}. 
If the graph $GL1$ representing a flat system's architecture is \emph{disconnected}, i.e. consists of $N$ graphs % 
that are either connected graphs or single vertices (we denote this set by $GN$), 
then we identify SCSs in each of these graphs % 
separately and could also parallelise these computations (cf. also example below). 

Thus, we start the service (de)composition by construction the set $GN$ from the system's architecture on level $L_{1}$. 
In the case $GL1$ is \emph{connected}, we simply have $GN = \{GL1\}$. 
Then for each graph we proceed as follows:

\noindent
(1) We identify DT services and delete these vertices from the graph. 
This is non-recursive procedure with time complexity $O(n)$, where $n$ is a number of vertices.

\noindent
(2) We apply the recursive elimination technique \emph{One-Way-Catch-Them-Young} (OWCTY, \cite{OWCTY}) to remove the LT services and
the reversed OWCTY technique to remove the TT services.

\noindent
(3) The removed services become single services on the level $L_{2}$. 
To the rest of the graph, we apply one of the existing  algorithms for the 
decomposition into SCCs (FB~\cite{idetifyingSSCs}, OBF~\cite{OBF}, etc.).

To illustrate these ideas, we continue with the example 
from previous section.  
We represent the system from Fig.~\ref{fig:example_comm2}  as a directed graph to apply a decomposition algorithm, cf. Fig.~\ref{fig:graph}.
This graph is disconnected and consists of two connected graphs, $G_{1}$ and $G_{2}$, therefore we can analyse them in parallel.
For $G_{1}$ we need to perform the following steps:
 
\noindent
(1) First of all, we identify  $A_{12}$ as a DT service, delete it from the graph (system's subservice $S_{1}$ on the level $L_{2}$).

\begin{figure}[ht!]
  \begin{center}
   \includegraphics[width=9cm, natwidth=499,natheight=277]{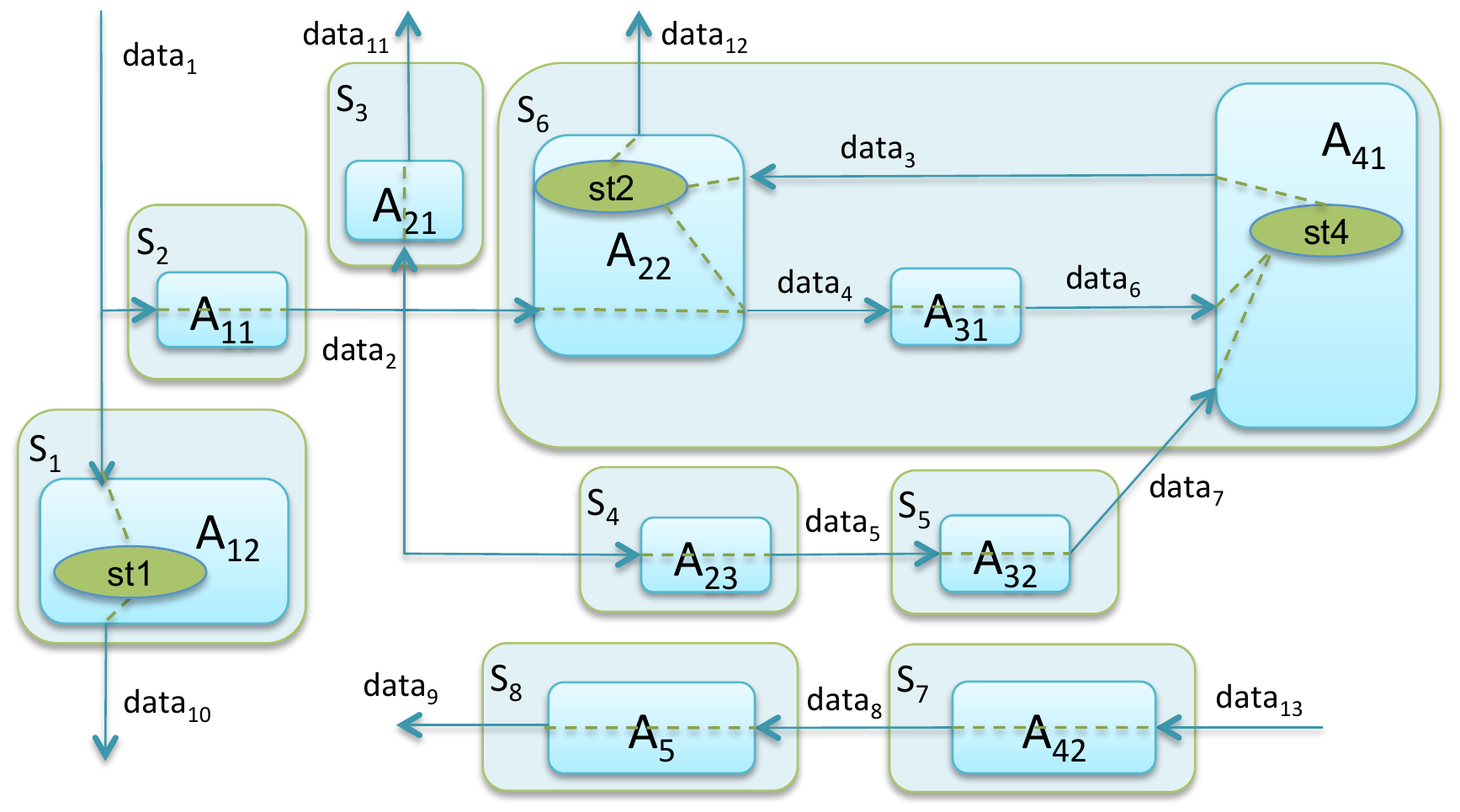}
    \caption{% 
    Architecture of $S$ (level $L_{2}$)}
    \label{fig:L2a}
  \end{center}
\end{figure}

\noindent
(2) Then, we apply the OWCTY elimination techniques to identify and remove the LT and TT services: 
	\begin{itemize}
	\item
	On the first run of the algorithm, we identify $A_{11}$ and $A_{21}$ as an LT and an TT services respectively and delete them.  
	They become $S_{2}$ and $S_{3}$ on the level $L_{2}$. 
        \item 
        After this, $A_{23}$ (which is commonly trivial in $G_{1}$) becomes an LT service and should be deleted, becoming $S_{4}$ on $L_{2}$. 
        	Then, the same situation will be with $A_{32}$: it should be deleted as an LT service and becomes $S_{5}$. 
        \item 
        The rest of the graph (vertices $A_{22}$, $A_{31}$, $A_{41}$) with corresponding edges contains neither LT nor TT vertices. 
         This is  an input for the SCC-decomposition algorithm.
	\end{itemize}
(3)
In our example we apply a variant of the FB algorithm. 
The time complexity of this algorithm  is $O(n*(n+m))$, where $n$ is a number of vertices and $m$ is the number of edges in this graph. 
The FB algorithm identifies that the vertices $A_{22}$, $A_{31}$, $A_{41}$ build a single SCS, which becomes $S_{6}$ on the level $L_{2}$. 
$G_{2}$ has only two vertices, 
a leading trivial SCS $A_{42}$ and a terminating trivial SCS $A_{5}$. 
Thus, on the level $L_{2}$ it gives us two subservices of the system: $S_{7}$ and $S_{8}$. 
Fig.~\ref{fig:L2a} presents the result of the architecture optimisation.

%===========================================================
\section{Efficient Checking of Properties}
\label{sec:eff_check}
 
A property can be represented  by relations over data and control flows on the system's channels. 
Let for a relation $r$, $I_{r}$ and $O_{r}$ be the sets of input and output channels of the system used in this relation.
For each channel from $O_{r}$ we recursively compute all the sets of the dependent input channels. 
Their union, restricted to the input channels of the system (for the case we exclude properties' specification over local channels), 
$\idep_{r}$ should be equal to $I_{r}$, otherwise we should check whether the property was specified correctly 
(i.e. exclude (human) error in the specification of the property, e.g., when a wrong channel identifier was used) and if so, precise it or eliminate unnecessary constraints:

\noindent
(1) If a channel $x$ belongs to $\idep_{r}$ but not to $I_{r}$, we need to extend $r$ explicitly specifying that this property should hold for any values on $x$.

\noindent
(2) If a channel $x$ belongs to $I_{r}$ but not to $\idep_{r}$, this means that $r$ is unnecessary strict, because it contains assumptions on irrelevant input channel.

From $O_{r}$ we obtain the set $OutComp_{r}$ of services having these channels as outputs, compute the union of corresponding sets $Sources^{L_{2}}$. 
This union together with $OutComp_{r}$ give us  the minimal part of the system needed to check the property $r$.  
This allowed us, especially in the case of cloud-supported processing, 
to reduce the costs of monitoring, testing or verification.

Assume we have to check relation $r_1(data_{10}, data_{13})$ specified for the system presented on Fig.~\ref{fig:L2a}. 
It is easy to see, that there is no dependency between $data_{13}$ and $data_{10}$. We have here $I_{r_{1}} = \{data_{13} \}$ and $O_{r_{1}} = \{ data_{10} \}$. 
$data_{10}$ is a single output channel of $S_{1}$, i.e. $ \outstreams(S_{1}) = \{ data_{10} \}$. This gives us $OutComp_{r_{1}} = S_{1}$. 
Thus, $\idep(S_{1}, data_{10}) = \{data_{1}\}$ which is already equal to $\idep_{r_{1}}$, because $data_{1}$ is a system input. 
First of all we need to exclude error in the specification of the property, e.g. the case where $r_1$ is meant to be specified over $data_{10}$ and $data_{1}$.
If $r_{1}$  was specified without such errors, we need 
(1) to extend it explicitly specifying that it should hold for any values on $data_{1}$, and 
(2) eliminate unnecessary constraints on $data_{13}$ from $r_{1}$.
Because $Sources^{L_{2}}(S_{1}) = \emptyset$, we need only the service $S_{1}$ to check this property.
 
Assume now we have to check relation $r_2(data_{1}, data_{12})$ specified for the same system.
For this case $I_{r_{2}} = \{data_{1} \}$ and $O_{r_{2}} = \{ data_{12} \}$. 
\\
$data_{12}$ is a single output channel of $S_{6}$, % 
and therefore $OutComp_{r_{2}} = S_{6}$. 
\\
$\idep(S_{6}, data_{12}) = \{data_{2}, data_{7}\}$ (both channels are local), $Sources^{L_{2}}(S_{6}) = \{ S_{2}, S_{5}\}$. 
\\
$data_{2} \in \outstreams(S_{2})$, which gives us $\idep(S_{2}, data_{2}) = \{data_{1}\}$ (this is a system input channel, therefore it should be a part of $\idep_{r_{2}}$). 
$Sources^{L_{2}}(S_{2}) = \emptyset$.
\\
$data_{7} \in \outstreams(S_{5})$, $\idep(S_{5}, data_{7}) = \{data_{5}\}$ (also local), $Sources^{L_{2}}(S_{5}) = \{ S_{4}\}$. \\
$data_{5} \in \outstreams(S_{4})$, $\idep(S_{4}, data_{5}) = \{data_{2}\}$ (local, cf. above), $Sources^{L_{2}}(S_{4}) = \emptyset$. \\
$\idep_{r_{2}} = \{ data_{1} \} = I_{r_{2}}$, there is no need to extend the property or eliminate some to its constraints. 
To check this property we need the following set of services:\\
$OutComp_{r_{2}} \cup Sources^{L_{2}}(S_{6}) \cup Sources^{L_{2}}(S_{2})  \cup Sources^{L_{2}}(S_{5})  \cup  Sources^{L_{2}}(S_{4}) = $\\ 
$\{  S_{2},  S_{4}, S_{5}, S_{6} \} $.

 %=================================================================================
\section{Remote Computation}  
\label{sec:remote}

In the previous section we have analysed,  which part of the overall information about the system and its input data 
is necessary  to check the corresponding property. 
In the similar way, we can trace the influence of the system environment's properties on the properties of the system itself. 
However, the decision, 
which information need to be processed locally (local services) and which need to be sent to the cloud (remote services), 
should be made not only according to this analysis of a (logical) system architecture, but also taking into account the following aspects of data flows between services:

\noindent
(1) measure for costs of the data transfer/ upload to the cloud \emph{UplSize(f)}:
 size of  messages (data packages) within a data flow $f$ and  frequency they are produced. 
This measure can be defined on the level of logical modelling, 
where we already know the general type of the data and 
can also analyse the corresponding  service (or environment) model to estimate the frequency the data are produced;

\noindent
(2) measure 
for requirement of using high-performance computing and cloud virtual machines, \emph{Perf(X)}: 
complexity of the computation within a service $X$, which can be estimated on the  level of logical modelling as well.

On this basis, we build a system architecture, optimised for remote computation. % 
We associate an  \emph{UplSize} measure to each channel (to each data flow), and 
an  \emph{Perf} measure to each elementary service in the system (i.e. for any service on the abstraction level $L_{1}$). 
A limit, above which  a remote computation is desired, is denoted by \emph{HighPerf}. 
A limit of the  \emph{UplSize} measure, above which a limited up-/download is desired, is denoted by \emph{HighLoad}, 
i.e. if the source of the data is deployed locally then the receiving services should preferably also deployed locally,  and 
i.e. if the source is deployed in the cloud then the receiving services should preferably also deployed in the cloud. 
 We do not take into account costs of the transfer of software service themselves, 
assuming that this aspect is already covered by the \emph{Perf} measure. 
The \emph{UplSize} measure should be analysed only for the channels that aren't local for the services on abstraction level $L_{2}$. 

Using graphical representation, we denote channels with \emph{UplSize} $\gg$ \emph{HighLoad} by {\color{Red} thick red} arrows, and 
the services with  \emph{Perf} $\gg$ \emph{HighPerf} by  \textbf{white} % 
colour, 
where all other channel and services are marked  {\color{CornflowerBlue} blue}. 
We use the same colouring notation when building the corresponding tables for the values above limits as well as when represented a system by the corresponding directed graph. 
For the compressed table representation, which allows readable representation of the large systems' measures, 
we omit the concrete values of the measures leaving only the notes on relations to the defined limits. 
While measuring \emph{Perf} on the level $L_{1}$, we mark a service  by a sum of \emph{Perf} measures of its subservices on the level $L_{2}$. 
Thus, if the \emph{Perf} measure of at least one subservice is high, the same hold for the service's \emph{Perf}. 

This analysis can also be done automatically, 
 provided all the measures \emph{UplSize} and \emph{Perf}  are defined for data and control flows as well as for (elementary) services. 
 Using our approach, we can prove a number of  properties for service composition according to these measures.
We start with the architecture from abstraction level $L_{2}$ and represent it as a directed graph. % 
Then we simply remove edges that correspond to channels with low \emph{UplSize}, and obtain a set of connected graphs. 
If any two vertices (or graphs) are ``connected'' by a single edge that is ``split'' and has no source vertex (this represents a situation where a system input goes to many systems's services), 
we see them as a connected graph too.
Each of the  graphs becomes a service on the abstraction level $L_{3}$ represents a system architecture, optimised for remote computation.

 \begin{figure}[ht!]
  \begin{center}
   \includegraphics[width=9cm, natwidth=377,natheight=139]{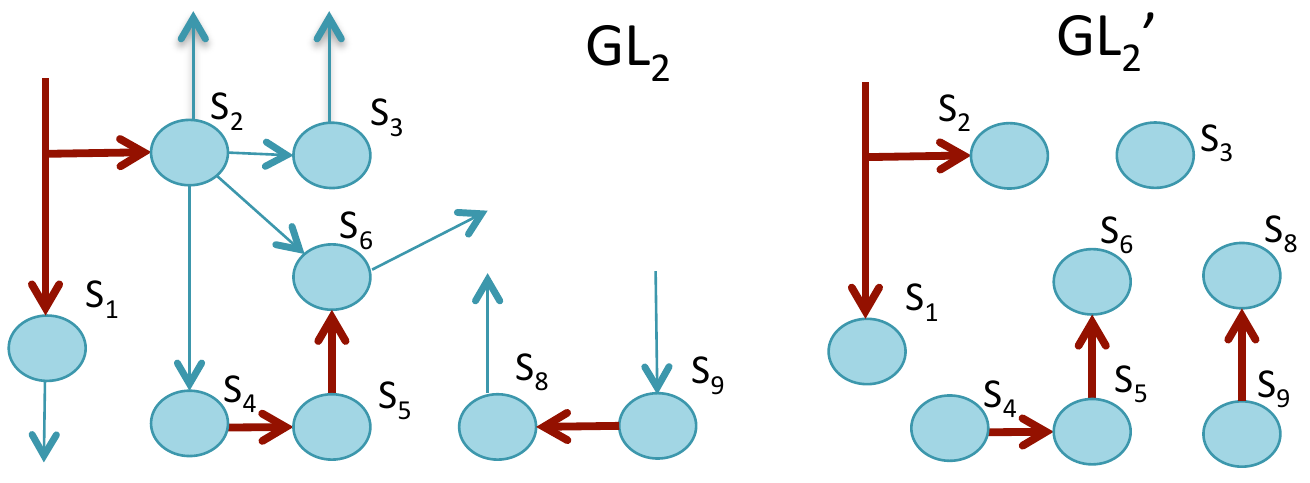}
    \caption{% 
   Service (de)composition according to the channel measures \emph{UplSize}}
    \label{fig:graph2}
  \end{center}
\end{figure}  

\noindent
Assume the following  case for the system $S$ from the previous examples 
(cf. also Fig.~\ref{fig:L2a}): 
\begin{itemize}
\item
channels with \emph{UplSize} value $\gg$ \emph{HighLoad}: \\
$data_{1}$, $data_{4}$, $data_{5}$, $data_{6}$, $data_{7}$, $data_{8}$.
\item
services with \emph{Perf} value $\gg$ \emph{HighPerf}: 
$A_{22}$, $A_{23}$, $A_{41}$, $A_{42}$, and on the abstraction level $L_{2}$: $S_{2}$, $S_{4}$, $S_{6}$, $S_{7}$.
\end{itemize}
 $data_{4}$ and $data_{6}$ are local channels of $S_{6}$, their  \emph{UplSize} measure is not relevant for this case. 
We represent the system from the abstraction level $L_{2}$ as a directed graph $GL_{2}$ (cf.  Fig.~\ref{fig:graph2}). 
After removing edges that correspond to channels with low \emph{UplSize}, we obtain a set $GL_{2}'$ of connected graphs.

\begin{figure}[ht!]
  \begin{center}
   \includegraphics[width=9cm, natwidth=498,natheight=279]{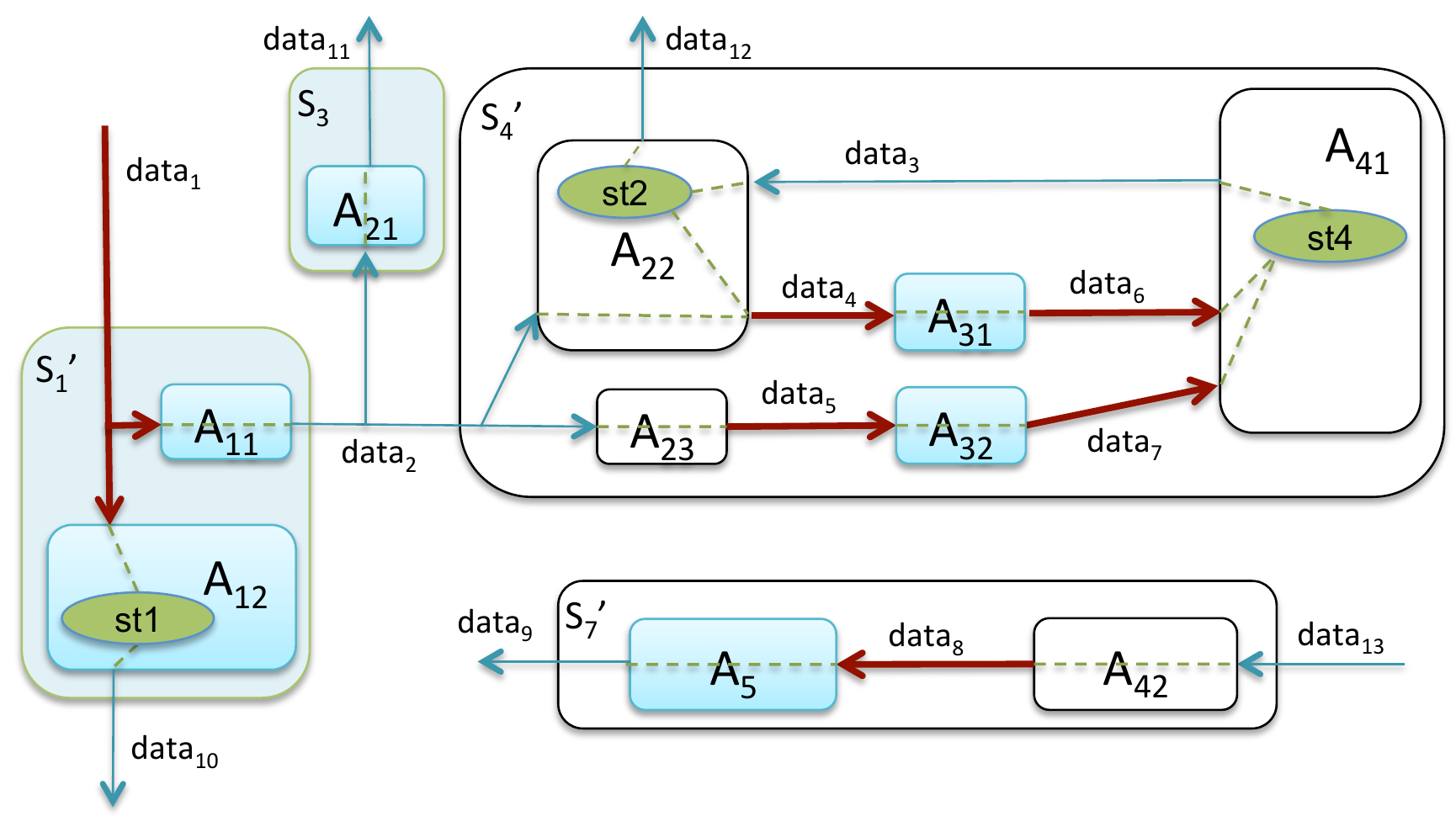}
    \caption{%  
    Optimised architecture of $S$  (Level $L_{3}$)}
    \label{fig:remote}
  \end{center}
\end{figure} 

\noindent
Fig.~\ref{fig:remote} represents a system architecture, optimised for remote computation. 
On the abstraction level $L_{3}$ (cf. Fig.~\ref{fig:remote}), $S_{1}$ and $S_{2}$ are composed together into a new service $S_{1}'$: $data_{1}$ corresponds to a data flow, where messages (data packages) 
 have large size and come with a high frequency, but  $data_{2}$ has very low frequency (e.g.,  
 $S_{2}$ realise a filtering function according to a given criteria), therefore it make more sense to deploy $S_{1}$ locally, together with $S_{2}$.

There are no changes for service $S_{3}$, but $S_{4}$, $S_{5}$, and $S_{6}$ are composed into $S_{4}'$, and $S_{7}$ with $S_{8}$ into $S_{7}'$. 
The services $S_{4}'$ and $S_{7}'$ have  \emph{Perf} measure higher \emph{HighPerf}, therefore using high-performance computing and cloud virtual machines is required for these services.

\section{Semi-Automatic Formal Verification}
\label{sec:verify}

To support this approach on verification level, we have also build a set of corresponding theories in Higher-Order Logic (HOL)
 using an interactive semi-automatic theorem prover (proof assistant) Isabelle~\cite{npw}.  
This proof assistant is based on polymorphic HOL extended with axiomatic type classes, and support the proof of arbitrary mathematical theorems in interactive manner. 
Proofs are constructed in the structured proof language Isar \cite{Isar}. 
The advantage of this proof language is that the proofs  are easy readable for both human and machine, which is not the case for many proof languages.

\begin{figure}[ht!]
 \centering% 
   \includegraphics[width=7cm, natwidth=546,natheight=315]{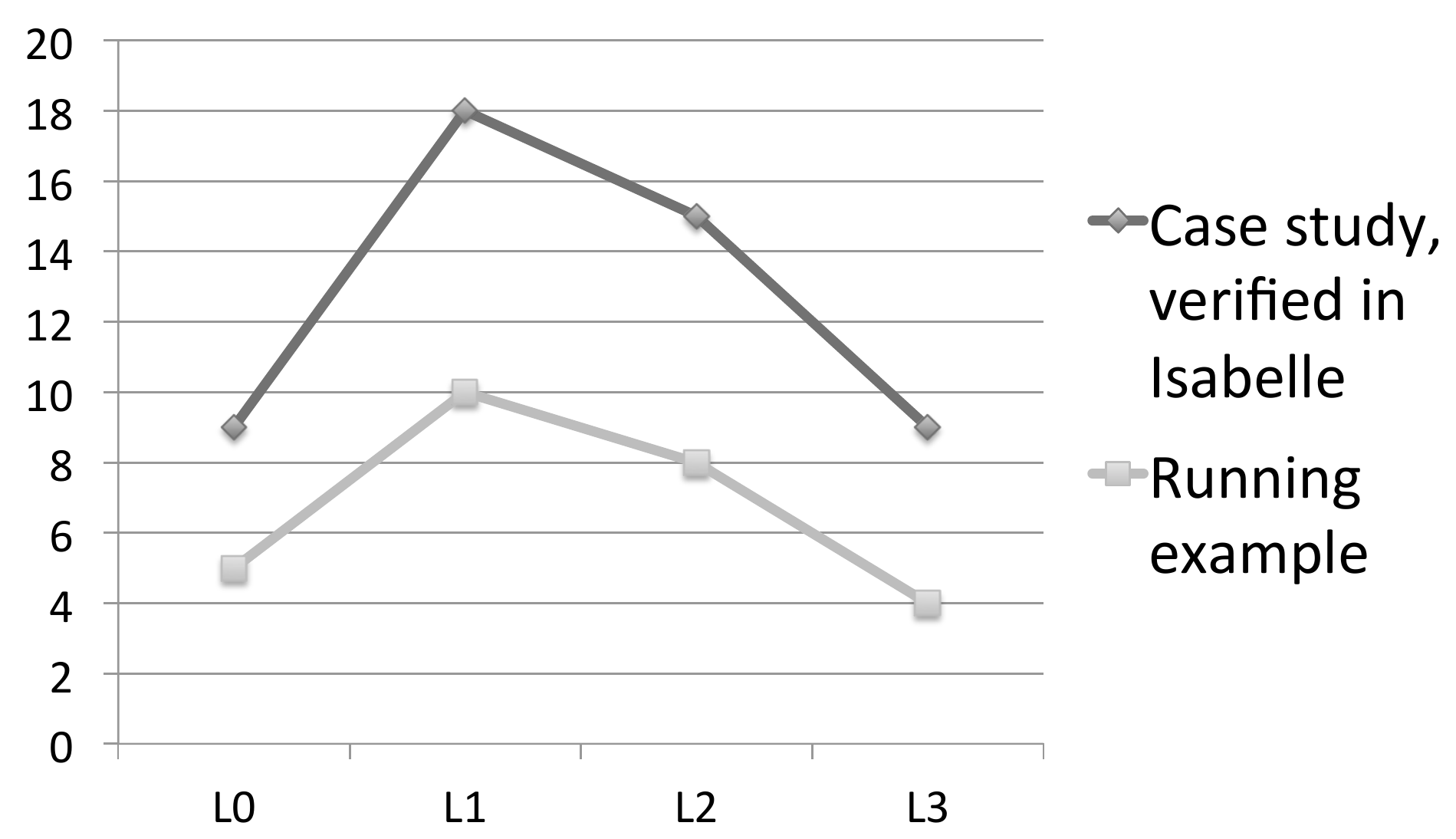}% 
    \caption{Correlation between the number of services and abstraction levels: Case study an the running example (cf. Section \ref{sec:example})}
    \label{fig:caseStudy}
\end{figure}

The representation of our approach in Isabelle/HOL contains 
approx. 70 ge\-ne\-ral axioms and lemmas \cite{CompDependencies-AFP}, which are necessary to analyse 
 dependencies within a system using  Isabelle in (semi-)automatic way, e.g.\ 
to verify whether a given set of services is equal to the set of (in)direct  sources of some service, or whether the set of services is equal to the minimal set needed to check a certain property.  
To show a feasibility of  the approach, we also have done a case study using our  formalisation. 
In this case study \cite{CompDependencies-AFP}, we started with 9 services connected with 24 channels on the abstraction level $L_{0}$ (among them 4 services are specified using local variables). 
Fig. \ref{fig:caseStudy} presents how the number of services was changed among the abstraction levels. %

~\\
Overall, the case study contains more than 300 lemmas, approx. 50\% of them can be composed and proven automatically, using the predefined schema, which is a part of our formalisation. 
All the technical details of formalisation within the theorem prover Isabelle as well as the corresponding proofs for a case study to elaborate the approach are presented in~\cite{CompDependencies-AFP}. 
Approx. 90\% of the proofs (general as well as for a concrete system from the case study) 
are constructed using Isabelle's component Sledgehammer~\cite{Sledgehammer}.  
Sledgehammer discharges proof goals applying to them resolution based
first-order automatic theorem provers (ATPs) 
and satisfiability modulo theories (SMT), which makes the human-related part of the proof simpler and faster.
\section{Discussion and Related Work}
\label{sec:related}
 
 The approach presented above extends our previous work in \cite{dentum_tb,spichkova_tb_decomp,spichkova2011decomp}  
 by providing  
 $(i)$ formal approach that allows 
 modelling of (data  and control flow) dependencies between components, 
 and  $(ii)$   associated algorithm for architecture optimisation towards efficient verification, testing, and monitoring of system's properties.  
We leave here out of scope the automatisation of the construction of 
  the functions $\idep$, $\odep$, etc. for a concrete case. 
  For the example presented in this paper as well as for the case study, verified in Isabelle, these functions were specified manually. 
Thus, this automatisation is planned for the future work.
However,
 one of the advantages of our current work is formal semi-automatic analysis of $(i)$ system architecture, and $(ii)$  
 the sufficiency of the set of system's components to check a certain property. 
 Thus, in our work we touch the following research areas: 
modelling communication between components, system decomposition, as well as  architecture modelling. 
In the rest of the section we discuss the related work on them.

\emph{Modelling communication:} 
Various languages and techniques have been proposed to represent
communication between components/processes, for example, Bergstra's
Algebra of Communicating Processes \cite{ACP}, Hoare's approach on
Communicating Sequential Processes \cite{CSP} and its extension
\cite{hilderink2003}.  Magee et al. \cite{Magee:1995:SDS:645385.651497} tried to
combine the ideas of operational semantics with Milner's $\pi$
calculus, calculus of mobile processes
(cf. \cite{Milner:1982:CCS:539036,DBLP:books/daglib/0098267}).  Reo, a channel-based coordination
model for component composition, represents a co-algebraic view on
this area \cite{Reo04,Meng2012799}. 
The work presented in~\cite{Vogel-Heuser_IECON} defines an extensive support to the components communication and time requirements, while the model discussed in~\cite{IEEE_INDIN_2011} proposes general ideas on model for distributed automation systems. 
\\
\indent
Our approach, in contrast, is focusing not on the communication in general but on the dependency aspects and how their analysis can be used to increase the efficiency of properties checking.
 
\emph{System decomposition.} 
There is a large number of approaches in the area of systems
decomposition (see, e.g., \cite{PR99,TUM-I0818}).  % 
 In general we can  say, that decomposition in many cases leads to a
refinement of a system, where by composing a system from components we
can implicitly build a new level of abstraction of system
representation (cf. \cite{broy_refinement,spichkova2008refinement}).
\\
\indent
The main difference and
the main contribution of our current work, also compared with our previous work
on  formal decomposition~\cite{spichkova2011decomp}, is $(i)$ an
 extension of the specification approach to associate
subspecifications with subservices or architectural components in a
more usable, readable way,  $(ii)$
 focusing on the aspects essential for the efficient checking of system's properties, % 
 $(iii)$ taking into account aspects that are important 
for the case of using high-performance computing and cloud virtual machines.  
To this end, our focus in the current paper goes
also beyond ideas presented in~\cite{spichkova2011decomp}, where
readability and manageability of specification were discussed in relation
to inconsistencies and incomplete specifications.

\emph{Architecture  modelling.}  
An introductory overview of foundations and applications of the model
driven architecture can be found in~\cite{DBLP:conf/ecmdafa/2006}.
There is a large collection of approaches on architecture elaboration,
with different aims and domain orientations, e.g., 
Medvidovic and  Taylor \cite{Medvidovic2000} introduce a classification and
comparison framework for software architecture description
languages, Malek et al. \cite{DBLP:journals/tse/MalekMM12}
presents a framework for improving distributed system's architecture
and their deployment,
 Broy et al. \cite{broy_janus} focus on specification and design of services
and layered architectures.
A number of approaches, e.g., \cite{ADdemistif},  propose  several meta-models % 
that introduce relationships between architectural design  decision
alternatives and activities related to them.  
A number of architecture description languages have been
developed to specify compositional views of a system on an abstract
level, e.g., TrustME~\cite{Schmidt2001trustByContracts}, which
combines software architecture specification approaches with ideas of
design-by-contract and allows capturing of  
behavioural interaction patterns 
between large-scale components of software and systems architectures.
\\
\indent
Our current work is focused on modelling of a logical service-oriented architecture. The main focus of our approach is on 
elaborating for a concrete system, which part of the system is sufficient to check a certain property, and how to optimise the logical architecture  towards efficient monitoring, testing, and verification of system's properties, also for the case of remote connection. 
% 

%================================================
 
\section{Conclusion and Future Work}
\label{sec:conclusions}

In this paper, we presented a formal approach to modelling and analysis data and control flow dependencies between components.  
We described the theory which lies behind the approach and presented a running example to illustrate the main ideas of the approach.  
We conclude that our approach allows 
\begin{itemize}
\item 
to specify the dependencies within a system formally, 
\item 
to elaborate for a concrete system, which part of the system is sufficient to check a certain property,
\item 
 to analyse tracing and reliability aspects,
\item
to optimise the architecture of a given system, also taking into account such aspects as 
$(i)$ costs of data transfer/ upload to the cloud, 
and $(ii)$ requirements of using high-performance computing and cloud virtual machines,
\end{itemize}
These results are especially important for analysis and optimisation of remotely deployed distributed control systems: 
our approach  allows system decomposition oriented towards efficient checking of system
properties -- 
it allows to send  to or from the cloud only the information really needed for the monitoring,
 testing or verification of the properties of interest. 
 
 Another contribution of our approach is a semi-automatic support of the presented ideas on verification level, 
  an interactive semi-automatic theorem prover  Isabelle/HOL.
 
 \emph{Future Work:}  
In the future work, we intent to automatise the construction of 
  the functions that specify the dependencies ($\idep$, $\odep$, etc.) for a concrete case on each level of abstraction. 
One of the other possible directions of our future work is on combination of the ideas presented above with our previous work on 
analysis of crypto-based components \cite{CryptoBasedCompositionalProperties-AFP}, to analyse the data dependencies between components wrt. secrecy/security  properties.
 
\bibliographystyle{IEEEtran}

\end{document}